\documentclass[twocolumn,floatfix]{aastex701}%This is used for arXiv submission

\newcommand{\fermi}{\textit{Fermi}}
\newcommand{\swift}{\textit{Swift}}
\newcommand{\nustar}{NuSTAR}
\newcommand{\g}{$\gamma$-ray}

\usepackage{subfigure}
\usepackage{amsmath,multirow, url,hyperref}
\usepackage[normalem]{ulem}
\usepackage{booktabs}
\usepackage{subcaption}
\newcommand{\revision}{\color{black}}
\newcommand{\production}{\color{black}}
\newcommand{\hide}[1]{}

\shorttitle{RIAF Interpretations for \swift{} and \nustar{} Observations of the LLAGN NGC 4278}
\shortauthors{}
\graphicspath{{./}{figures/}}
\begin{document}

\title{Interpreting \swift{} and \nustar{} Observations of the Low-Luminosity Active Galactic Nucleus NGC 4278 with Radiatively Inefficient Accretion Flows and Implications for Neutrino Emission}

\author[orcid=0000-0003-2916-0211]{Abhishek Das}
\affiliation{Department of Physics, The Pennsylvania State University, University Park, PA 16802, USA}
%\affiliation{Department of Astronomy \& Astrophysics, The Pennsylvania State University, University Park, PA 16802, USA}
\affiliation{Institute for Gravitation and the Cosmos, The Pennsylvania State University, University Park, PA 16802, USA}
\email{ajd6518@psu.edu}

\author[orcid=0000-0001-6674-4238]{Qi Feng}
\affiliation{Department of Physics and Astronomy, University of Utah, Salt Lake City, UT 84112, USA}
\email{u6053889@utah.edu}

\author[]{Eleanor Young}
\affiliation{Department of Physics and Astronomy, University of Utah, Salt Lake City, UT 84112, USA}
\email{} 

\author[orcid=0000-0003-3820-0887]{Ashwani Pandey}
\affiliation{Department of Physics and Astronomy, University of Utah, Salt Lake City, UT 84112, USA}
\email{} 

\author[orcid=0000-0003-2579-7266]{Shigeo S. Kimura} 
\affiliation{Frontier Research Institute for Interdisciplinary Sciences, Tohoku University, Sendai 980-8578, Japan}
\affiliation{Astronomical Institute, Tohoku University, Sendai 980-8578, Japan}
\email{shigeo@astr.tohoku.ac.jp}

\author[orcid=0000-0002-5358-5642]{Kohta Murase}
\affiliation{Department of Physics, The Pennsylvania State University, University Park, PA 16802, USA}
\affiliation{Department of Astronomy \& Astrophysics, The Pennsylvania State University, University Park, PA 16802, USA}
\affiliation{Institute for Gravitation and the Cosmos, The Pennsylvania State University, University Park, PA 16802, USA}
\affiliation{Center for Gravitational Physics and Quantum Information, Yukawa Institute for Theoretical Physics, Kyoto University, Kyoto 606-8502, Japan}
\email{murase@psu.edu}

\correspondingauthor{Abhishek Das}
\email{ajd6518@psu.edu}
\correspondingauthor{Qi Feng}
\email{qi.feng@utah.edu}

%% Mark off the abstract in the ``abstract'' environment. 
\begin{abstract}

We report the first \nustar{} hard X-ray observations of the low-luminosity active galactic nucleus NGC 4278. The source is clearly detected beyond 10 keV with a hard X-ray spectrum consistent with a power law of photon index between $2.2$ and $2.5$ without evidence for a high-energy cutoff. The X-ray flux is low compared to the active state in 2021, but exhibits variability by a factor of $\sim2$ on a timescale of a month. We discuss the origin of the hard X-ray emission and explore its connection to gamma rays and high-energy neutrinos. We explain the X-ray data, including both quiescent and active states, using a radiatively inefficient accretion flow (RIAF) model with a variable accretion rate.
We also show that TeV gamma rays cannot escape from the RIAF disk, and very high-energy gamma rays observed by LHAASO are likely to originate from outer regions such as jets and winds, which is consistent with our results favoring a magnetically arrested disk state of the RIAF disk. We also discuss hidden neutrino emission from RIAFs together with possible connections to coronae of active galactic nuclei with standard, radiatively efficient disks. 
\end{abstract}

%% Keywords should appear after the \end{abstract} command. 
%% The AAS Journals now uses Unified Astronomy Thesaurus concepts:
%% https://astrothesaurus.org
%% You will be asked to selected these concepts during the submission process
%% but this old "keyword" functionality is maintained in case authors want
%% to include these concepts in their preprints.
\keywords{\uat{High energy astrophysics}{739} --- \uat{Low luminosity active galactic nuclei}{2033} --- \uat{LINER galaxies}{925} --- \uat{X-ray astronomy}{1810} --- \uat{Accretion}{14}  --- \uat{Gamma-ray astronomy}{628}}

%% From the front matter, we move on to the body of the paper.
%% Sections are demarcated by \section and \subsection, respectively.
%% Observe the use of the LaTeX \label
%% command after the \subsection to give a symbolic KEY to the
%% subsection for cross-referencing in a \ref command.
%% You can use LaTeX's \ref and \label commands to keep track of
%% cross-references to sections, equations, tables, and figures.
%% That way, if you change the order of any elements, LaTeX will
%% automatically renumber them.
%%
%% We recommend that authors also use the natbib \citep
%% and \citet commands to identify citations.  The citations are
%% tied to the reference list via symbolic KEYs. The KEY corresponds
%% to the KEY in the \bibitem in the reference list below. 

\section{Introduction} \label{sec:intro}

Ground-based gamma-ray observatories have discovered more than 200 sources of TeV gamma-ray radiation, which can only be produced in some of the most extreme environments in the Universe. New types of TeV gamma-ray sources are still being discovered, e.g., recurring novae \citep{HESS2022}. 
Ground-based water Cherenkov telescopes, such as LHAASO and HAWC, have the unique advantage of a large field of view (FoV) and a high duty cycle (not affected by the Sun or the Moon) for discoveries of new TeV sources. 

The first catalog of the highest-energy (from 1 TeV to above 100 TeV) gamma-ray sources detected by LHAASO \citep{Cao2024} includes 32 new TeV sources. Only one of these sources was identified by the LHAASO Collaboration as a candidate extragalactic source, 1LHAASO J1219+2915, which is coincident with the low-luminosity active galactic nucleus (LLAGN) NGC 4278. It was only detected by the WCDA (1--25 TeV) of LHAASO at a statistical significance of $\sim7\sigma$, with a power-law index of 2.67$\pm$0.17 and as a point source (the 95\% UL on its 30\% containment radius is 0.08$^\circ$),
Subsequently, LHAASO reported a $\sim5$-month period when NGC 4278 exhibited an active state with elevated TeV gamma-ray flux, between August 23, 2021, and January 10, 2022 \citep{Cao2024_NGC4278}. 

%There are three nearby known TeV sources: 1ES 1218+304 (0.94$^\circ$ away from 1LHAASO J1219+2915), 1ES 1215+303 (0.97$^\circ$ away), and W Comae (1.09$^\circ$ away). Considering the 95\% localization uncertainty of 0.09$^\circ$ of 1LHAASO J1219+2915, it is much more likely to be associated with NGC 4278, located at 0.05$^\circ$ away, than any of the known TeV sources. 

NGC 4278 is a nearby ($z=0.0021$) low-ionization nuclear emission-line region (LINER)/LLAGN at a luminosity distance of $\sim$16.1 Mpc, exhibiting a two-sided parsec-scale ``S-shaped'' jet structure from radio observations \citep{Giroletti2005, Tremblay2016}. The radio jets in the north and south directions show asymmetry in their apparent velocity, which can be explained by either a mildly relativistic jet ($\beta\sim0.75$) at a small viewing angle (2$^\circ$--4$^\circ$) or jets at larger viewing angle strongly interacting with the surrounding medium \citep{Giroletti2005}. 
An intriguing transient GeV gamma-ray source, 1FLT J1219+2907, spatially coincident with NGC 4278, was detected at $\sim5\sigma$ by \fermi{}-LAT in 2009 between March 5 and April 5, with a 95\% error radius of 0.5$^\circ$ \citep{Baldini2021}. 
In the X-ray band, NGC 4278 has been observed by XMM-Newton in 2004 and Chandra between 2005 and 2010, revealing X-ray flux variability on timescales of years \citep{Pellegrini2012} and months \citep{Younes2010}. Evidence for flux variation by $\sim10\%$ on timescales of hours has also been reported \cite{Younes2010}. The source has not been observed in X-rays after 2010, except for a short \swift{} observation ($\sim$0.9 ks exposure) in 2021.  

In this work, we describe the first \nustar{} observations of NGC 4278, the measured X-ray spectrum beyond 10 keV, and its variability; we interpret the X-ray results using a radiatively inefficient accretion flow model; we also discuss the possible association between the X-ray emission and the observed TeV gamma rays, as well as the hypothetical neutrinos.

%%%%%%%%%
% Observations and data analysis
%%%%%%%%%
\section{Observations and Data Analysis}
\label{sec:obs_data}
%

%%%%%%%%%
\subsection{\nustar{}}
\label{subsec:nustar}
%%%%%%%%%
The Nuclear Spectroscopic Telescope Array (\nustar{}) uses two co-aligned grazing incidence telescopes with two independent focal plane modules (FPMA and FPMB) to detect hard X-rays between 3 and 79 keV \citep{Harrison_2013}. \nustar{} covers a $13' \times 13'$ FoV and achieves an on-axis point spread function of 18" \citep{Harrison_2013}. 

\nustar{} observed NGC 4278 for the first time on December 6, 2024, for $\sim$43.1 ks (observation ID 61002012002). Subsequently, \nustar{} observed the source again on January 13, 2025, for $\sim$45.7 ks (observation ID 61002012004). 

The \nustar{} data were analyzed using standard software (\texttt{NuSTARDAS}) as part of \texttt{HEASOFT}. An absorbed power-law model with a cross-normalization parameter was used to simultaneously fit the spectrum from each of the two observations. The total Galactic neutral hydrogen column density is $N_H=2.22 \times 10^{20} \textrm{ atoms cm}^{-2}$ toward the direction of NGC 4278 \citep{Willingale13}. We performed the fit to the X-ray spectra with both $N_H$ free and $N_H$ fixed at the Galactic value. 

%%%%%%%%%
\subsection{\swift{}}
\label{subsec:swift}
%%%%%%%%%

The X-Ray Telescope (XRT) on the Neil Gehrels {\it Swift} Observatory is capable of detecting X-rays from $\sim$0.3~keV to 10~keV \citep{Gehrels04, Burrows05}. 

Previously, \swift{}-XRT observed the vicinity of NGC 4278 two times, including a 923-s exposure on November 28, 2021 (observation ID 03109562002), during the LHAASO-reported active state. 
Following the announcement of a TeV gamma-ray detection by LHAASO, \swift{} performed 10 observations of NGC 4278 between May 2024 and January 2025. These more recent observations include a 1584-s exposure on January 13, 2025, taken on the same day as one of the two \nustar{} observations.

%obsid 00089850002 on January 13, 2025, simultaneous with the NuSTAR observations, exposure 1584 s

The \swift{}-XRT data were analyzed following the standard procedure using \texttt{HEASOFT} and modeled using an absorbed power law. A joint fit with both \swift{}-XRT and \nustar{} data taken on January 13, 2025, was also performed. 

\begin{figure}[hb]
\centering
\includegraphics[width=1.0\linewidth]{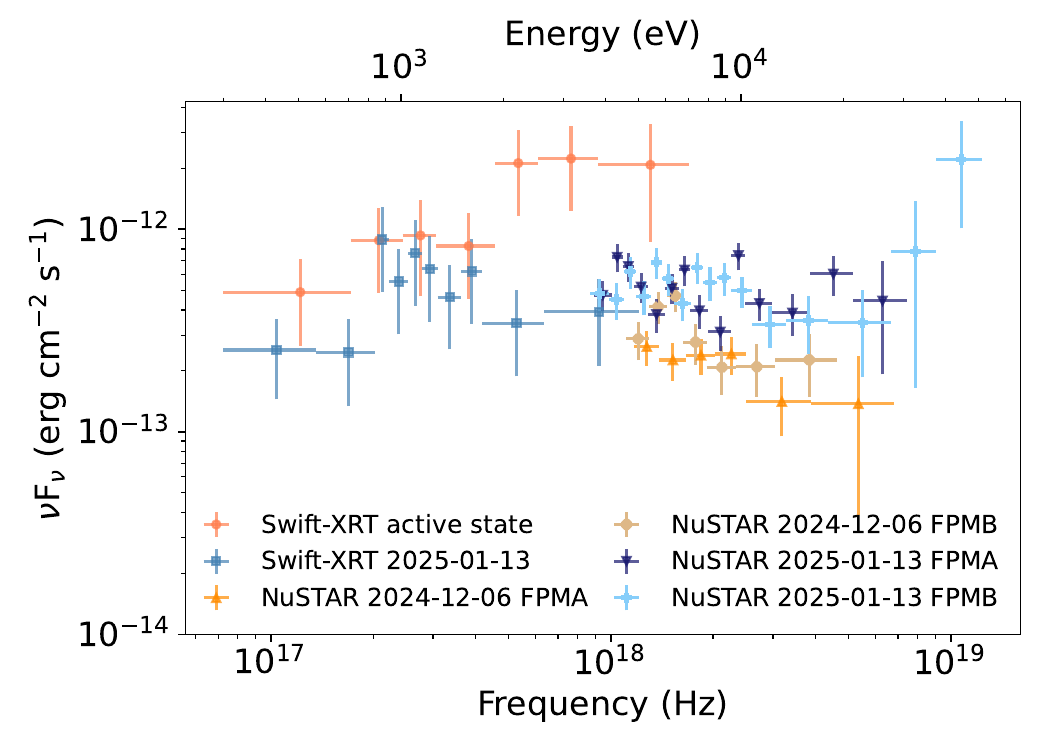}
\caption{The X-ray spectra of NGC 4278. Quasi-simultaneous spectra measured by \swift{}-XRT (blue squares) and \nustar{} (dark blue down triangles for FPMA and light blue plus signs for FPMB) on January 13, 2025, show a moderate flux level. The \swift{}-XRT spectrum measured during the LHAASO-reported active state (pink circles) shows an X-ray active state that is more prominent above 2 keV. The \nustar{} spectrum (orange up triangles for FPMA and gold cross signs for FPMB) measured on December 6, 2024, shows a low-flux state. The flux points are corrected for the Galactic neutral hydrogen absorption.}
\label{fig:xspec}
\end{figure}

%%%%%%%%%
% Results
%%%%%%%%%
\section{Results}
\label{sec:res}

%
%%%%%%%%%
\subsection{X-ray spectrum}
\label{subsec:xspec}
%%%%%%%%%

Fig.~\ref{fig:xspec} shows the X-ray spectrum measured quasi-simultaneously by \swift{}-XRT and \nustar{} on January 13, 2025. The \swift{}-XRT spectrum measured on November 28, 2021, during the LHAASO-reported active state, as well as the first \nustar{} spectrum measured on December 6, 2024, are shown for comparison.

\begin{table}[]
\centering
    \caption{Best-fit parameters of absorbed power-law models for the \swift{}-XRT and \nustar{} X-ray spectrum measured on January 13, 2025.}
    \label{tab:xspec}
\begin{tabular}{ c | c | c }
\hline\hline
\multirow{2}{*}{Parameter} & \multicolumn{2}{c}{Best-Fit Value and 1$\sigma$ Uncertainty}          \\%\hline
 & $N_H$ fixed & $N_H$ free \\ \hline
Index & $2.02\pm0.05$ & $2.25\pm0.07$  \\ \hline
Normalization  & \multirow{2}{*}{$3.2\pm0.3$} & \multirow{2}{*}{$5.0\pm0.6$}  \\ 
($10^{-4} \mathrm{keV}^{-1} \mathrm{cm}^{-2}$) & &  \\ \hline
$N_H$ & \multirow{2}{*}{$2.22$}   & \multirow{2}{*}{$24.9 \pm 7.7$}  \\
($10^{20} \textrm{ atoms cm}^{-2}$)&    &    \\ \hline
$\chi^2$/DOF & $125.6/89$ & $104.1/88$  \\
\hline
\end{tabular}
\end{table}

The X-ray spectrum measured by \swift{}-XRT during the active state was hard, with a power-law photon index of $1.4\pm0.3$, suggesting that the synchrotron emission in the spectral energy distribution (SED) continues to rise beyond a few keV. The active-state X-ray spectrum shows a much larger increase in the flux above 2 keV compared to the flux below 2 keV (see Fig.~\ref{fig:xspec}). 

The recent \nustar{} observations measured an almost flat X-ray spectrum without any sign of a high-energy cutoff. The two \nustar{} observations showed that the 3-30 keV X-ray spectrum became marginally harder, with the power-law photon index changing from $2.5\pm0.2$ on December 6, 2024, to $2.2\pm0.1$ on January 13, 2025 ($N_H$ fixed at the Galactic value).

The measured X-ray spectrum is curved toward low energies after correcting for the Galactic neutral hydrogen absorption. This suggests either an intrinsic curvature, or the existence of additional absorption beyond the Galactic neutral hydrogen if the intrinsic spectrum follows a power law from 0.3 keV to a few tens of keV. 
The joint \swift{}-XRT and \nustar{} X-ray spectra measured on January 13, 2025, are fitted to an absorbed power-law model (\texttt{tbabs*po}) with the neutral hydrogen column density either fixed to the Galactic value or left free (see Table~\ref{tab:xspec}). 
A simple $F$-test shows that leaving $N_H$ free significantly improves the power-law fit at a $p$-value of $5\times10^{-5}$ (or $\sim 4\sigma$). Although the curvature can also be intrinsic and could be explained by a radiatively inefficient accretion flow (see Section~\ref{subsec:riaf}). 

A relativistic reflection disk-corona model has been tested on the X-ray spectrum; however, due to limited statistics, the parameters are not well constrained. 

It is worth noting that the \nustar{} spectrum measured on January 13, 2025, becomes background-dominated above 20 keV, and the flux measurements in the two high-energy bins above 30 keV from FPMB are not statistically significant (light blue pluses in Fig.~\ref{fig:xspec}). 

%%%%%%%%%
\subsection{X-ray flux variability}
\label{subsec:var}
\begin{figure}[hb]
\centering
\includegraphics[width=1.0\linewidth]{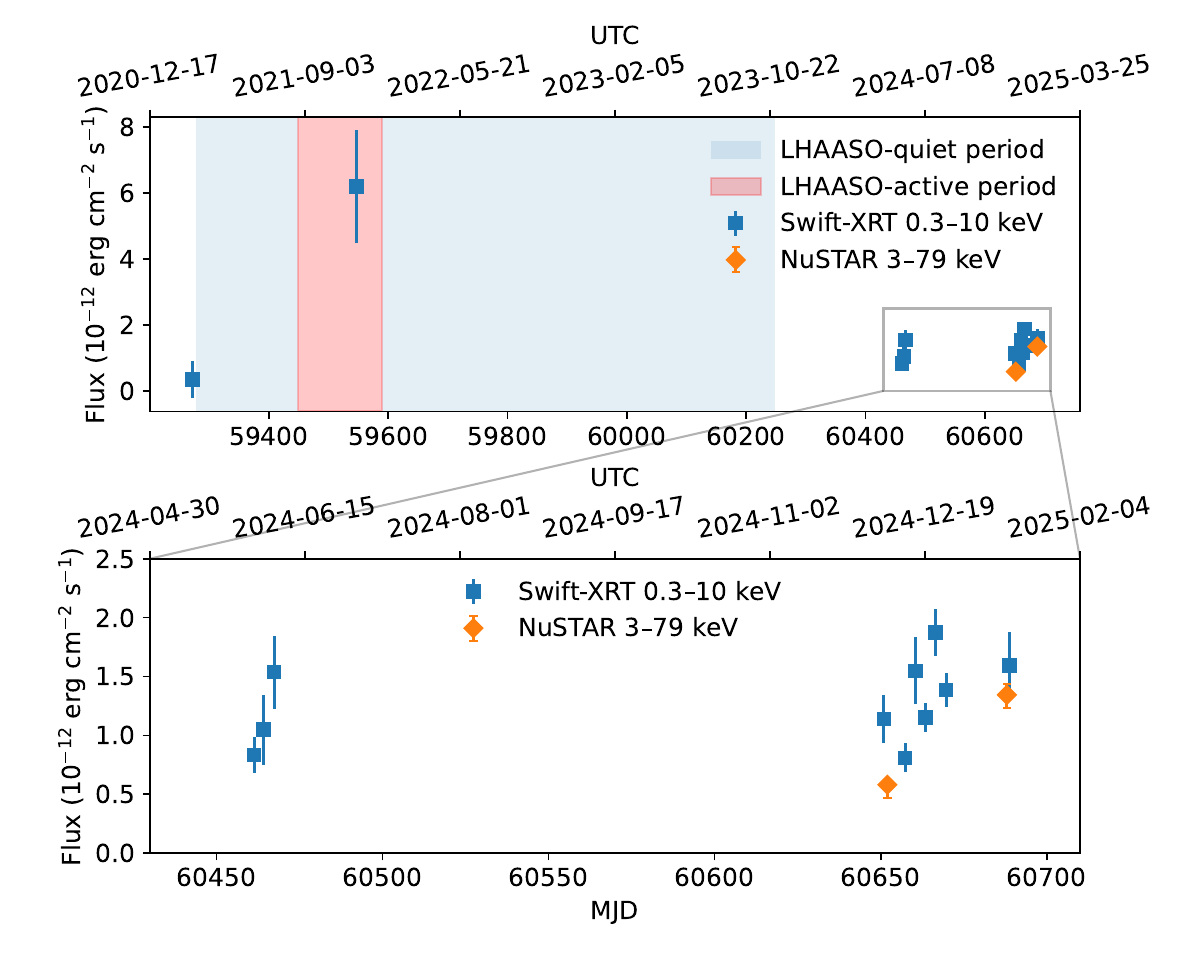}
\caption{\textit{(Top)} The X-ray light curve of NGC 4278 between 2020 and 2025, showing an active X-ray flux state during the LHAASO-reported active periods. \textit{(Bottom)} A zoomed-in view of the variability during the X-ray campaign in 2024 and 2025, showing a low to moderate flux state with variability by a factor of $\sim2$ over timescales of weeks to months. }
\label{fig:xlc}
\end{figure}

%%%%%%%%%
Despite the sparsity of X-ray monitoring before 2024, the serendipitous \swift{}-XRT observation during the LHAASO-reported ``active'' state on November 28, 2021 captured an elevated 0.3-10 keV X-ray flux of $(6.2\pm1.8) \times 10^{-12}\, \mathrm{erg\,cm^{-2}\,s^{-1}}$, as shown in Fig.~\ref{fig:xlc} (\textit{top}). 

The 10 recent \swift{}-XRT observations between May 2024 and January 2025 showed a much lower 0.3-10 keV X-ray flux, with a mean and standard deviation of $(1.29\pm0.35) \times 10^{-12}\, \mathrm{erg\,cm^{-2}\,s^{-1}}$, as shown in Fig.~\ref{fig:xlc} (\textit{bottom}). Comparing the low and high X-ray fluxes, the \swift{}-XRT measured 0.3-10 keV flux during the 2021 gamma-ray active period likely corresponds to an X-ray active state, showing a higher flux by a factor of $\sim5$ with a statistical significance of $\sim2.7 \sigma$. 

Although the X-ray flux was lower during the 2024 -- 2025 campaign, it was still variable on timescales of days to months. A constant fit to the low-state 0.3-10 keV flux yielded a $\chi^2=34.56$ over 9 degrees of freedom, corresponding to a $p$-value of $\sim7\times10^{-5}$, suggesting variability was observed over a few months. The 0.3-10 keV flux increased by almost a factor of 2 from December 2024 to January 2025. The fractional variability \cite{Vaughan03} from the 10 low-state XRT observations is $0.21\pm0.06$. 

The two \nustar{} observations also showed a factor of $\sim2$ increase in the 3-79 keV hard X-ray flux, from $(5.8^{+1.0}_{-1.2}) \times 10^{-13}\, \mathrm{erg\,cm^{-2}\,s^{-1}}$ on December 6, 2024, to $(1.4\pm0.1) \times 10^{-12}\, \mathrm{erg\,cm^{-2}\,s^{-1}}$ on January 13, 2025. We therefore refer to the \nustar{} observation on December 6, 2024, as the ``quiescent'' flux state, and that on January 13, 2025, as the ``moderate'' flux state. 

%%%%%%%%%
\subsection{Broadband SED}
\label{subsec:SED}
%%%%%%%%%

\begin{figure}[b]
\centering
\includegraphics[width=1.0\linewidth]{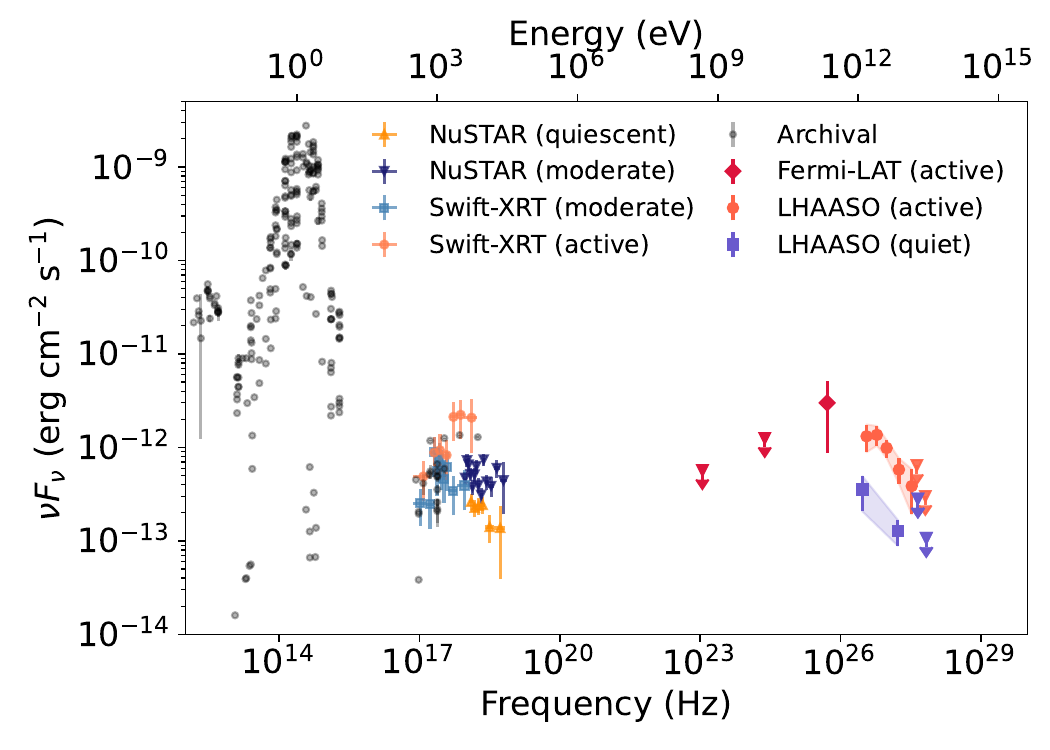}
\caption{The broadband SED of NGC 4278. The active state SED is measured by \swift{}-XRT (pink circles), \fermi{}-LAT (red diamonds) \citep{Bronzini2024}, and LHAASO (tomato red circles) \citep{Cao2024_NGC4278}. The ``quiescent'' \nustar{} SED was measured on December 6, 2024. The ``moderate'' X-ray SED was measured quasi-simultaneously by \swift{}-XRT and \nustar{} on January 13, 2025 (see Sec.~\ref{subsec:xspec} and \ref{subsec:var}). The archival SEDs are taken from the ASI Space Science Data Center (SSDC) SED Builder \citep{Stratta2011}.}
\label{fig:sed}
\end{figure}

%Investigate Fermi-LAT UL; and IR/MM SED, refer to \cite{Kimura2021} for expected SED from RIAF
%\setcounter{footnote}{0}

Fig.~\ref{fig:sed} shows the broadband SED of NGC 4278 during the quiescent and moderate X-ray flux states in 2024 and 2025, together with the soft X-ray SED during the active state in 2021, as well as the archival SEDs across many wavelengths. The TeV gamma-ray spectra measured by LHAASO \citep{Cao2024_NGC4278} during the active state (April 10, 2021 -- August 28, 2021) and the quiet state (March 5, 2021 -- October 31, 2023, excluding the active period) are shown. NGC 4278 is nominally not detected by \fermi{}-LAT, but \cite{Bronzini2024} reported a $4.3\sigma$ detection in an analysis of the \fermi{}-LAT data from March 2021 to October 2022. We adopted the \fermi{}-LAT spectrum during the active state of the source from \cite{Bronzini2024}. 
The archival SEDs are taken from the ASI Space Science Data Center (SSDC) SED Builder \footnote{ASI Space Science Data Center (SSDC) SED Builder:
\url{https://tools.ssdc.asi.it/SED/}} \citep{Stratta2011}. The archival data shown are from multiwavelength surveys from radio to X-ray energies \citep{2016A&A...588A.103B,
1999A&A...349..389V,
2010AJ....140.1868W,
1990IRASF.C......0M,
1994yCat.2125....0J,
1998AJ....115.1693C,
1996ApJS..103..427G,
1997ApJ...475..479W,
1992ApJS...79..331W,
1970ApJS...20....1D,
2003MNRAS.341....1M,
2007MNRAS.376..371J,
2007ApJS..171...61H}.

\section{Theoretical Modeling of SEDs}\label{sec:discuss}
\subsection{RIAF Model}
\label{subsec:riaf}
\begin{figure*}
    \centering
    \includegraphics[width=\linewidth]{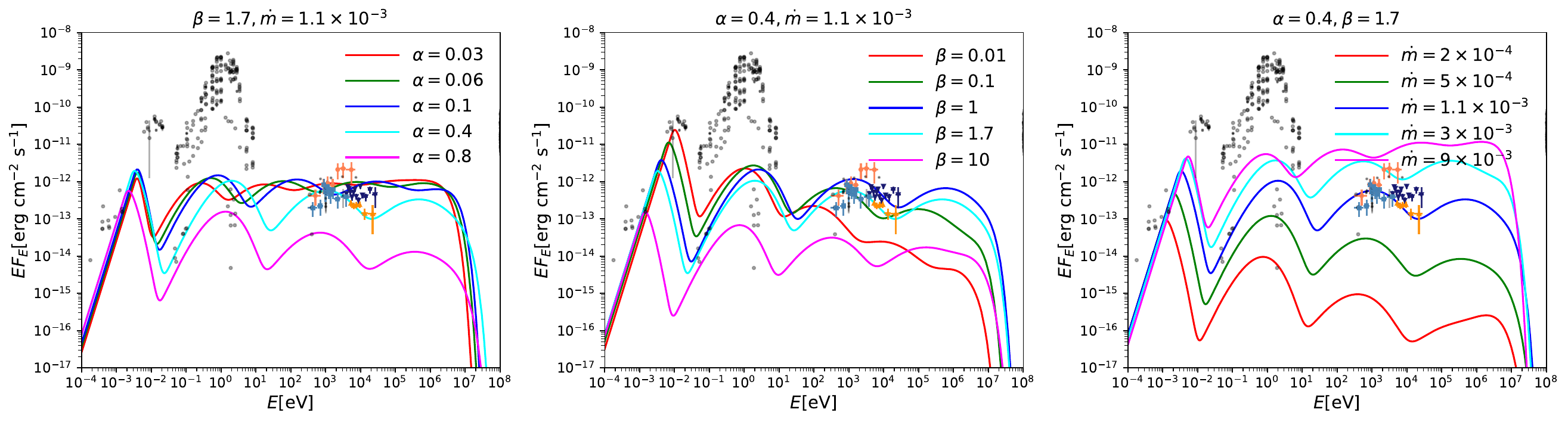}
    \caption{Left: SEDs with varying $\alpha$ for constant values of $\beta$ and $\dot{m}$. Middle: Same with $\beta$ for constant values of $\alpha$ and $\dot{m}$. Right: Same with $\dot{m}$ for constant values of $\alpha$ and $\beta$. The archival and X-ray data points have the same legends as Fig. \ref{fig:sed}. The purple and black data points are from \cite{Nemmen1} and \cite{Banerjee2019}, respectively.}
    \label{fig:param-var}
\end{figure*}
We attempt to explain the SED with a radiatively inefficient accretion flow (RIAF) model, where thermal electrons emit multi-wavelength photons via synchrotron and Comptonization processes \citep{NarayanADAF,BlandfordBegelman,Mahadevan:1996jf,Pesce:2021adg}. Such a model explains keV-MeV photons without the need for a big blue bump, which is typically absent in LLAGNs. The model is characterized primarily by three parameters: the viscosity parameter, $\alpha$, the pressure ratio of gas to magnetic fields, $\beta$, and the normalized accretion rate, $\dot{m}$, which is related to the accretion rate $\dot{M}$ as $\dot{m} = \dot{M}c^2/L_{\text{Edd}}$. We also define the normalized radius $\mathcal{R}=R/R_S$. The radial velocity, number density, proton thermal temperature, magnetic field, Thomson optical depth, and Alfvén velocity for the RIAF model are defined and analytically approximated as follows~\citep{Kimura2019, Kimura2021}:
\begin{eqnarray}
V_R &=& \alpha V_K/2 \simeq 3.4 \times 10^8~\mathcal{R}^{-1/2}_1 \alpha_{-1}~\rm cm/s, \\
n_p&=&\frac{\dot{M}}{4\pi m_p RHV_R}\nonumber\\
    &\simeq& 1.5 \times 10^7~\mathcal{R}^{-3/2}_1\alpha^{-1}_{-1}M^{-1}_{8.5}\dot{m}_{-3}~\rm cm^{-3},\\
%    &\simeq& 4.6 \times 10^8~\mathcal{R}^{-3/2}_1\alpha^{-1}_{-1}M^{-1}_8\dot{m}_{-2}~\rm cm^{-3},\\
k_BT_p&\approx&\frac{GMm_p}{4R}\simeq12~\mathcal{R}^{-1}_1~\rm MeV,\\
    B&=&\sqrt{\frac{8\pi n_pk_BT_p}{\beta}} \nonumber\\
    &\simeq& 84~\mathcal{R}^{-5/4}_1\alpha^{-1/2}_{-1}M^{-1/2}_{8.5}\dot{m}^{1/2}_{-3}\beta^{-1/2} {\rm G},\\
    %&\simeq& 1.5\times10^2~\mathcal{R}^{-5/4}_1\alpha^{-1/2}_{-1}M^{-1/2}_8\dot{m}^{1/2}_{-2}\beta^{-1/2}_1 {\rm G},\\
\tau_T&\approx& n_p\sigma_T R\simeq9.0\times10^{-3}~\mathcal{R}^{-1/2}_1\dot{m}_{-3}\alpha^{-1}_{-1},\\
%\tau_T&\approx& n_p\sigma_T R\simeq0.090~\mathcal{R}^{-1/2}_1\dot{m}_{-2}\alpha^{-1}_{-1},\\
\beta_A&=&\frac{B}{\sqrt{4\pi n_pm_pc^2}}\simeq0.16~\mathcal{R}^{-1/2}_1\beta^{-1/2},%\beta_A&=&\frac{B}{\sqrt{4\pi n_pm_pc^2}}\simeq0.050~\mathcal{R}^{-1/2}_1\beta^{-1/2}_1,
\end{eqnarray}
where $V_K=\sqrt{GM/R}$ is the Keplerian velocity, $H\approx R/2$ is the scale height, and $A_n = A/10^n$.

%The bolometric luminosity of the SED is given by
%\begin{equation}
%    L_\text{bol} \approx \eta_\text{rad,sd}\dot{m}_\text{crit}L_\text{Edd}(\dot{m}/\dot{m}_\text{crit})^2
%\end{equation}
%where $\dot{m}_\text{crit}\approx 0.03(\alpha/0.1)^2$ is the critical mass accretion rate above
%which RIAFs no longer exist.

The bolometric luminosity of the SED is computed with the given physical quantities in RIAFs, such as density, magnetic field, and electron temperature. We obtain the electron temperature by balancing the heating and cooling rates in the RIAF \citep[e.g.,][]{Kimura:2021ayq}. For the heating rate, we assume that half of the released gravitational energy is used for heating the plasma, and electrons receive $f_e$ of the released energy, i.e., the electron heating rate is given by $Q_e\approx f_e\dot M c^2/(2\mathcal R)$. We use the fitting formula given by \cite{Chael:2018aeq} for $f_e$. For cooling processes, we consider advection and radiative cooling. For radiative cooling, we take into account synchrotron, Comptonization, and bremsstrahlung processes (see \citealt{Kimura:2014jba} for details). First, we compute the electron temperature, $T_{e,\rm rad}$, by balancing radiative cooling and heating. If advection cooling is dominant, the temperature should be set to $T_{e,\rm adv}=f_eT_p$.  We use the lower electron temperature, $T_e={\rm min}(T_{e,\rm adv},~T_{e,\rm rad})$, to calculate the multi-wavelength spectrum.

In Fig.~\ref{fig:param-var}, we demonstrate the effects of varying each of the three main parameters. The left panel demonstrates the effect of different values of $\alpha$ on the SED for a constant $\beta$ and $\dot{m}$. We can see that for $\alpha\in[0.03,0.4]$, lowering the viscosity parameter $\alpha$ increases the ``Compton Y parameter" corresponding to a stronger Comptonization process, where photons on average gain more energy. 
%The black hole mass of NGC 4278 is set to $3\times{10}^8~M_\odot$ \citep{Wang2003}. 
Because of efficient Comptonization, the electron temperature in RIAF decreases, and thus, single scattering does not increase the photon energy significantly, but increases the optical depth, causing efficient multiple scatterings, resulting in a significant increase in Comptonization photons. On the other hand, a lower plasma $\beta$ results in a lower electron temperature because of efficient synchrotron cooling, thereby decreasing the Y parameter and causing inefficient Comptonization. The middle panel shows the effect of varying $\beta$ for constant $\alpha$ and $\dot{m}$. Finally, the last panel demonstrates the effect of $\dot{m}$ on the SED. More accretion results in increased Comptonization because of higher values of the Y parameter, resulting in greater flux at higher energies.

We present our results assuming a single emission zone. We show that the single-zone modeling is sufficient. For the multi-zone scenario, we consider three concentric emission regions with progressively larger radii, the smallest radius being $10 R_S$. The parameters $\alpha$ and $\beta$ are assigned to be constant across each emission region and behave exactly as they do in the single-zone scenario, while $\dot{m}$ at different radii is calculated as follows:
\begin{equation*}
    \dot{m}[i] = \dot{m}_0 \times (R[i]/R_0)^s
\end{equation*}
where $i = {0, 1, 2,...}$ is an index assigned to each concentric emission region, {\revision $\dot{m}_0$ and $R_0$ are the values corresponding to the innermost region,} $R$ is the emission radius and $s$ is a power-law index, which for our model is set to 0.5 \citep[see, e.g.,][]{Yuan_2012,Guo:2024gqc}.
%\citep[see, e.g.,][]{Yuan_2012, Yuan_2012_2, McKinney2012, Yuan_Narayan}. 
For the other parameters {\revision including $\dot{m}_0$}, we use the same values as for the single-zone model.

\begin{figure}%[ht]
\centering
\includegraphics[width=1.0\linewidth]{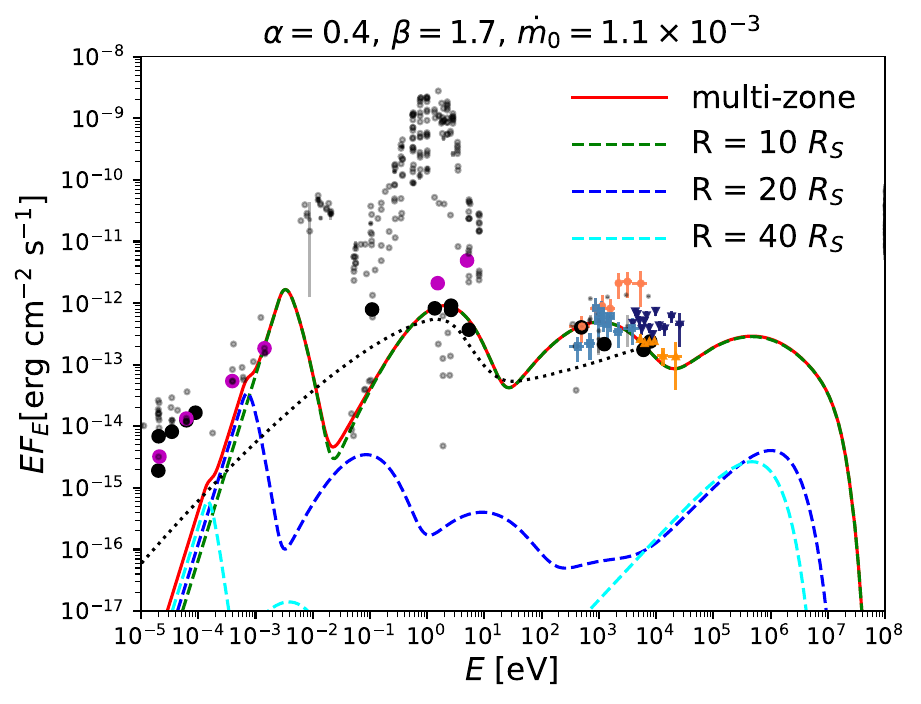}
\caption{Multi-zone SED from different radii with $R=(10-40)~R_S$, compared to a contribution from each radius. The archival and X-ray data points have the same legends as in Fig.~\ref{fig:sed}. 
The purple and black data points are %the same as in Fig. \ref{fig:param-var}.
from \cite{Nemmen1} and \cite{Banerjee2019} respectively. 
The LINER average SED is shown from \cite{Nemmen2} (black {\production dotted} line).
}
\label{fig:sed_mz}
\end{figure}

As seen in Fig.~\ref{fig:sed_mz}, the emission from the outer zones is several orders of magnitude smaller than that from the innermost region. 
This is because the synchrotron and Comptonization luminosities from RIAFs are very sensitive to their electron temperature \citep[see][]{Kimura2021}. In the outer zones, the electron temperatures are limited by advection cooling instead of radiative cooling, resulting in lower temperatures and causing weak radiation across the relevant wavelengths.
Although the outer zones may contribute more significantly in X-ray bands in situations where the free-free emission dominates \citep[e.g.,][for Sgr A*]{Yuan:2003dc}, it is insignificant in moderately accreting sources like NGC 4278.

\subsection{Flux Normalization}
The model has large parameter degeneracy when only X-ray data are explained. 
In order to avoid potential degeneracy in our choice of best-fit parameters for the RIAF SED, we attempt to normalize the SED using the sub-mm/far-IR and optical data in addition to the X-ray data obtained in this work. 

However, since most of the flux in the optical band originates from the stellar population of the galaxy, we have to extract a component that is consistent with point-source emission~\citep{Banerjee2019, Nemmen1}. Similarly, the IR excess in the archival data around 0.1 eV is due to emission from heated dust \citep{Tang_dust4278}. Note that the radio-frequency data points would be attributed to the jet and not the RIAF. 

We use the RIAF data from \cite{Banerjee2019} to normalize the {\revision optical} component of the {\revision quiescent-state} SED. 
The moderate state SED is then obtained by only increasing $\dot{m}$ to the value at which the SED provides a fit to the moderate state \nustar{} data or active state \swift{}-XRT data. The physical explanation for this is that both states describe the same fluid flow, but the rate of accretion is greater during the active state. 
%As a result, the ratio between the X-ray peaks is smaller than that between the optical peaks for this state, and the peak energy of the X-ray spectrum is lowered, as one would expect from the results in the third panel of Fig.~\ref{fig:param-var}. 
Note that \cite{Nemmen2} presents an averaged LINER SED, which appears quite different from pure RIAF SEDs commonly used in the literature. This difference is likely expected, because the data-driven SED template effectively averages over a heterogeneous population and may include substantial contributions from components beyond the RIAF itself (e.g., jets, truncated thin disks, and stellar contamination), which can wash out the characteristic spectral features of the RIAF SED model. As seen in Fig.~\ref{fig:sed_mz}, the data in the optical band from \cite{Nemmen1} largely overshoots this averaged SED, which could be caused by the stellar contamination. 

\subsection{Main Results}
We vary the viscosity parameter $\alpha$ in the range [0.03, 1], the plasma $\beta$ in [$10^{-3}$, 10], and the accretion rate $\dot{m}$ within the range [$10^{-5}$, $10^{-2}$]. We find the best-fit values for $\alpha$ and $\beta$ to be 0.4 and {\revision 1.7} respectively. These are typical values for the RIAF model. The best-fit value of $\dot{m}$ is found to be {\revision $1.1\times10^{-3}$} for the quiescent X-ray state and {\revision $1.5\times10^{-3}$} for the moderate state. 
The best-fit SED models are shown in Fig. \ref{fig:sed_sz} and the corresponding parameters are listed in Table \ref{tab:RIAF1Z}. 

RIAFs can be sorted into one of two categories based on the magnetic topology. One of these is the magnetically arrested disk (MAD) \citep{Narayan_MAD}, where the magnetic stresses are comparable to those required to impede the inflow (magnetic arrest), leading to a flux-saturated, highly variable accretion.
The other is the standard and normal evolution (SANE) \citep{MADvsSANE_sim}, where the magnetization is much weaker in the bulk of the flow, and the flux saturation is absent, and the accretion proceeds in the standard turbulence driven by the magnetorotational instability~\citep{RevModPhys.70.1}. While values of $\beta$ vary depending on regions, our choice of $\beta$ infers the MAD-like disk, $\beta\sim 0.1-10$ at the disk midplane \citep[e.g.,][]{Chael:2018aeq}, while for SANE, it is usually in the range $\beta\sim10-100$ \cite[e.g.,][]{Kimura:2018clk}.
%\citep{Kimura2019,Kimura_MAD}. 
Although the plasma $\beta$ itself is not enough to determine whether the flow is in the MAD or SANE, the MAD would be necessary to launch powerful jets \citep{TNM11a}, which is consistent with the indication of jets in NGC 4278 \citep{Giroletti2005,chen2026physicaloriginveryhighenergygamma}. The required jet power without protons is $\gtrsim {10}^{39}-10^{43}~{\rm erg}~{\rm s}^{-1}$, depending on models, which can be comparable to or smaller than our inferred values, $\dot{M}c^2\sim(2-4)\times{10}^{43}~{\rm erg}~{\rm s}^{-1}$.   

\begin{table}[]
\caption{SED parameters for the single-zone RIAF model with $R=10R_S$. }
\begin{tabular}{c|c|c|c}
\hline \hline
RIAF state & $\alpha$ & $\beta$ & $\dot{m}$ \\ \hline 
quiescent state  & 0.4      & 1.7     & $1.1\times10^{-3}$  \\ 
moderate state       & 0.4      & 1.7     & $1.5\times10^{-3}$   \\    \hline

\end{tabular}
\label{tab:RIAF1Z}
\end{table}

\section{Discussions}\label{sec:discuss}
%%%%%%%%%
\subsection{Inability of Very High-Energy Gamma Rays to Escape from the RIAF}
\label{subsec:GRassoc}
%%%%%%%%%
\begin{figure}%[ht]
\centering
\includegraphics[width=1.0\linewidth]{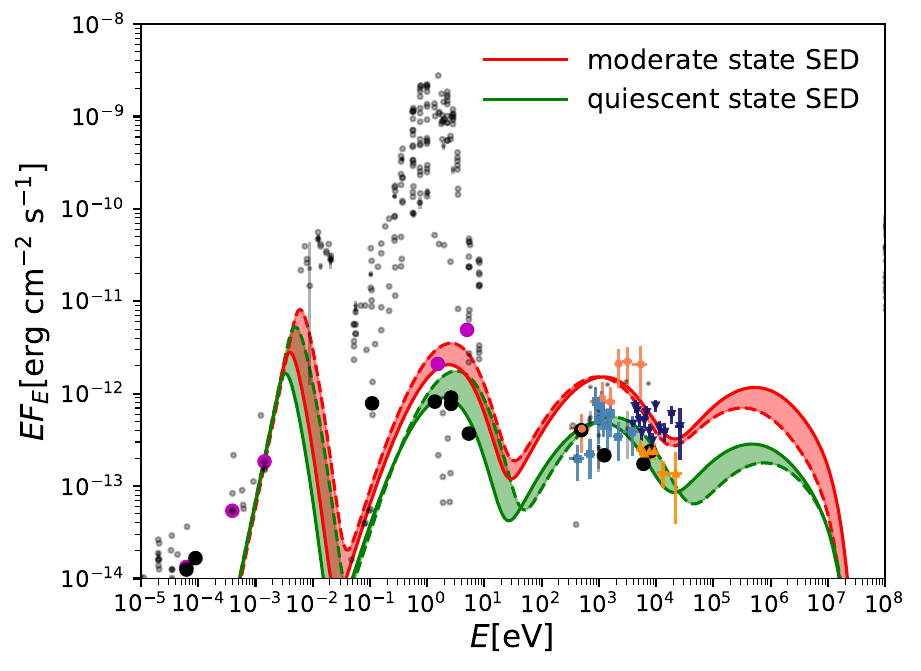}
\caption{{\production Bands representing the allowed range of SEDs}\hide{Best fit results } for a single-zone model using a typical emission radius of $10~R_S$. The best fit parameters for the SED {\production (shown using solid lines)} are listed in Table \ref{tab:RIAF1Z}. {\production The dashed lines represent the SEDs used by \cite{chen2026physicaloriginveryhighenergygamma} for the EIC scenario and correspond to $\alpha=0.6$ and $\beta=0.3$ with $\dot{m}=9\times10^{-4}$ for the quiescent state and $\dot{m}=1.8\times10^{-3}$ for the moderate state.} The archival and X-ray data points have the same legends as Fig.~\ref{fig:sed}. The purple and black data points are from \cite{Nemmen1} and \cite{Banerjee2019}, respectively.}
\label{fig:sed_sz}
\end{figure}

\begin{figure}%[ht]
\centering
\includegraphics[width=1.0\linewidth]{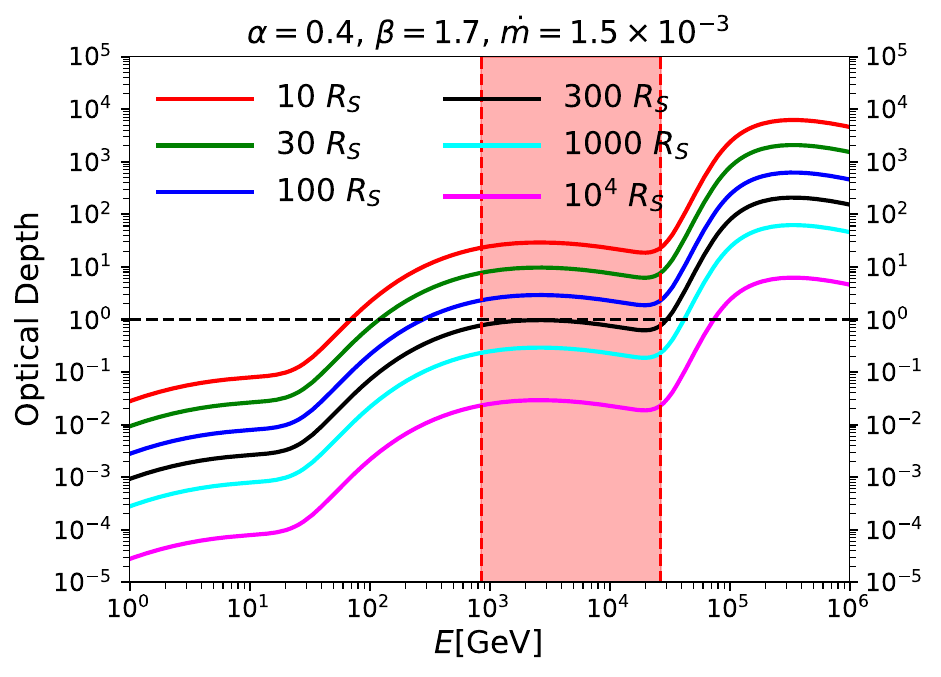}
\caption{Two-photon annihilation optical depths for the single-zone moderate state SED at different emission radii. The red shaded region is the LHAASO energy band.}
\label{fig:OD}
\end{figure}

It is interesting to ask whether the RIAF model explains very high-energy gamma rays from LLAGNs. For the RIAF model to explain the LHAASO data, the source must be transparent to gamma rays in the 1-25 TeV energy range. These gamma rays primarily interact with IR photons from the disk to produce electron-positron pairs. The optical depth for electron-positron pair production from the two-photon annihilation process is as follows \citep{Murase2022,Das_2024},
\begin{align}
   \tau_{\gamma \gamma \to e^+ e^-} &\approx 
    \eta_{\gamma \gamma} \sigma_T R \tilde{n}_{\rm disk} \left(\frac{\varepsilon_\gamma}{\tilde{\varepsilon}_\gamma^{e^+e^--\rm IR}}\right)^{\Gamma_{\rm disk} - 1} \nonumber \\ &\sim 30~\tilde{L}_{\rm disk, 39.7}\left(\frac{\varepsilon_{\rm IR}}{0.1~\rm eV}\right)^{-1}\left(\frac{R}{10~R_S}\right)^{-1}  \nonumber \\ &\quad {\left(\frac{M_{\rm BH}}{3\times10^8M_\odot}\right)}^{-1}\left(\frac{\varepsilon_\gamma}{\tilde{\varepsilon}_\gamma^{e^+e^--\rm IR}}\right)^{\Gamma_{\rm disk} - 1},
\end{align}
where $\eta_{\gamma\gamma}\sim 0.5$ is a coefficient that is dependent on the photon index of RIAF IR emission $\Gamma_{\rm disk}\sim1$~\citep{sve87,Murase:2015xka}, $\sigma_T \approx 6.65\times 10^{-25}\rm~cm^{2}$ is the Thomson cross section, $\tilde{L}_{\rm disk}\approx 4 \pi R^2 c  \tilde{n}_{\rm disk} \varepsilon_{\rm IR} \sim5\times{10}^{39}~{\rm erg}~{\rm s}^{-1}$ is the differential luminosity of the disk at $\varepsilon_{\rm IR}\sim0.1$~eV, $M_{\rm BH}$ is the black hole mass, $\sim3\times10^8M_\odot$ for our source \citep{Wang2003}, the characteristic energy of interacting gamma rays is $\tilde{\varepsilon}_\gamma^{e^+e^--\rm IR} \approx m_e^2 c^4 / \varepsilon_{\rm IR} \simeq 2.6\rm~TeV~(\varepsilon_{\rm IR}/0.1\rm~eV)^{-1}$, and the photon index of the disk spectrum is $\Gamma_{\rm disk}\sim1$. This equation is consistent with the numerical results shown in Fig. \ref{fig:OD}. Setting $\tau_{\gamma \gamma \to e^+ e^-}\leq 1$ in the LHAASO band to allow the gamma rays to escape gives us {\revision $R\gtrsim 300~R_S$}. This disfavors the RIAF disk itself as the potential region of the LHAASO gamma-ray emission (although the attenuation could be compensated by increasing the dissipation power). {\revision Furthermore, once photons approach the thermal energy threshold of $1-10$~MeV, Comptonization saturates, and Klein-Nishina-like effects suppress further energy gains \citep{Kimura2021}.} We therefore conclude that the {\revision GeV-}TeV gamma rays detected by {\revision{\it Fermi}-LAT and }LHAASO more likely originate from outer regions, such as the jet or wind driven by the LLAGN. 

A recent work from \cite{chen2026physicaloriginveryhighenergygamma} has demonstrated that the gamma-ray observations from LHAASO (both quiet and active states) can be explained by a jet modeled with an external inverse-Compton (EIC) scenario. They argue that the EIC process may be critical to very high-energy gamma-ray production in NGC 4278 and provides a better fit to the observed data than a synchrotron self-Compton (SSC) or leptohadronic model. The RIAF emission in this work can serve as target photons for the jet EIC process{\production , albeit with a slightly different set of parameters from those reported in Table \ref{tab:RIAF1Z} that results in a higher radio/optical/IR to X-ray flux ratio, as demonstrated in Fig. \ref{fig:sed_sz}}. As an alternative model, \cite{yuan2026tevgammarayslowluminosityactive} have explored the possibility of explaining the X-ray and gamma-ray states by leptohadronic radiation from a sub-relativistic wind.
%either by leptonic radiation from a moderately relativistic jet or by leptohadronic radiation from a sub-relativistic wind. 
However, these models cannot be distinguished with current observations. 

In the jet model, the gamma rays are more likely to originate from purely leptonic processes, although leptohadronic processes have also been considered \citep{chen2026physicaloriginveryhighenergygamma}. In the wind scenario, however, the gamma rays could have a primarily hadronic origin \citep{yuan2026tevgammarayslowluminosityactive}. In either case, the gamma-ray emitting region must be sufficiently far away from the center, {\revision $R\gtrsim 300~R_S$}, which is consistent with our conclusion.

\subsection{Low-luminosity AGNs as hidden neutrino sources and possible connections to turbulent coronae of Seyfert galaxies}
\begin{figure}%[ht]
\plotone{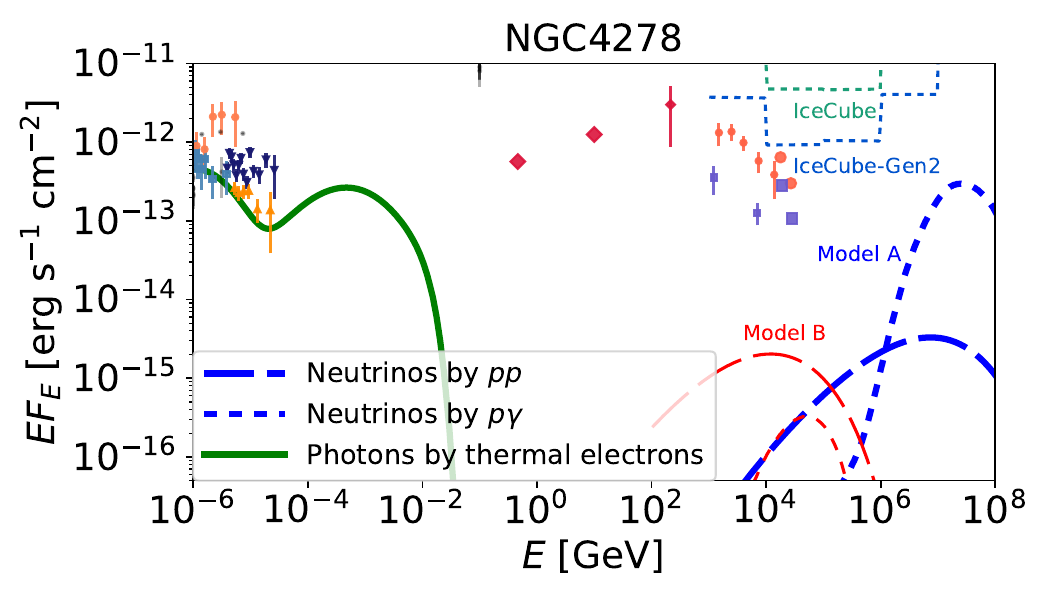}
\caption{Neutrino spectra from our RIAF model. The solid line is the photon spectrum emitted by thermal electrons {\revision corresponding to the quiescent state}. The dotted and dashed lines are neutrino spectra from $pp$ and $p\gamma$ interactions, respectively. The blue and red lines are for model A ($\eta_{\rm acc}=10$) and model B ($\eta_{\rm acc}=1.0\times10^4$), respectively (see text for the definition of $\eta_{\rm acc}$). {\revision The thin dotted lines are the 5$\sigma$ discovery potential for IceCube (orange) and IceCube-Gen2 (blue), respectively.}}
\label{fig:riaf_neu}
\end{figure}

As shown in the previous subsection, TeV gamma rays cannot escape from RIAFs. However, neutrinos freely escape from RIAF disks. It has been demonstrated that RIAFs in LLAGNs are capable of accelerating protons to PeV energies, leading to the production of PeV neutrinos through hadronuclear/photohadronic processes \citep{Kimura:2014jba,Kimura2019,Kimura:2020srn,Kimura2021}. Fig.~\ref{fig:riaf_neu} shows neutrino spectra from our RIAF models (see \cite{Kimura2019} for computational methods). Model A corresponds to efficient acceleration with $\eta_{\rm acc}=10$, where  $\eta_{\rm acc}$ is the acceleration efficiency defined as $t_{\rm acc}=\eta_{\rm acc} r_L/c$ with $t_{\rm acc}$ and $r_L$ being acceleration timescale and Larmor radius, respectively. Model B uses the lower acceleration efficiency with $\eta_{\rm acc}=1.0\times10^4$. Model A and Model B are similar to those in \cite{Kimura:2020srn} and \cite{Kimura2021}, respectively. {\revision We use the cosmic-ray injection efficiency of $\eta_{\rm CR}=0.05$ for both Models A and B}, which is defined as $L_{\rm CR}=\eta_{\rm CR}\dot Mc^2$ with $L_{\rm CR}$ being cosmic-ray luminosity. {\revision The power-law CR injection spectrum with a spectral index of 1 is assumed for simplicity.} 
We also stress that the turbulent RIAF model presented by \citet{Kimura2019} and \citet{Kimura2021} describes SANE-like and MAD-like ($\beta\lesssim 0.1-10$) disks in a unified manner, in the sense that the turbulent magnetic field is parameterized by the plasma beta in the same phenomenological framework.     

{\revision The maximum proton energy is obtained by equating $t_{\rm acc}$ to the shortest relevant timescale among diffusive escape, advection, cooling, and interaction timescales, which maintains the confinement condition $r_L\lesssim R$.} In Model A, cosmic-ray protons are accelerated to higher energies of $\sim100$ PeV, leading to efficient $p\gamma$ interaction and higher neutrino luminosity with $E_\nu\sim10$ PeV. On the other hand, the maximum energy of cosmic-ray protons in Model B is limited to $\sim1$ PeV, where the $pp$ channel is dominant, and pion production efficiency is lower, compared to Model A, leading to weaker neutrino emission. Since the magnetic field is not strong, pion synchrotron cooling is negligible in both Models. We should note that the cosmic-ray pressure in the RIAF is close to the theoretical limit, {\revision $P_{\rm CR}/P_{\rm th}\sim 0.5$ for both Models A and B~\citep{Murase:2020lnu}.}
{\revision Even with such a high CR pressure, the expected neutrino flux is lower than the $5\sigma$ discovery potential for future experiments, such as IceCube-Gen2 \citep{IceCubeGen2_2021}. Some experiments in the northern hemisphere would have larger effective area at $E_\nu\gtrsim 10^7$~GeV as they avoid the attenuation by Earth, the detailed examination of which would be left as future work.}
Although the proton-induced electromagnetic cascade leads to a broadband emission from keV X rays to sub-TeV gamma rays, this cascade emission is subdominant compared to the MeV gamma rays from the Comptonization process of thermal electrons~\citep{Kimura2019}.

%To roughly estimate a possible neutrino flux, we assume the TeV gamma rays from NGC 4278 detected by LHAASO entirely originate from relativistic protons. Since the LHAASO spectrum does not show evidence for any absorption, we assume the expected neutrino flux is comparable to the TeV gamma-ray flux. 
The expected neutrino flux in our model can be comparable to the TeV gamma-ray flux detected by LHAASO, although the RIAF disk should be ``hidden" in TeV gamma rays~\citep{Murase:2015xka}. This flux can be similar to the neutrino flux for the lepto-hadronic wind model discussed in \cite{yuan2026tevgammarayslowluminosityactive}. However, their wind model assumes cosmic-ray luminosity of $L_{\rm CR}\sim10^{44}\rm~erg~s^{-1}$, which would be larger than energy budget of the system, $\dot Mc^2\sim 4\times10^{43}~{\rm~erg~s}^{-1}\ll L_{\rm Edd}$, as estimated by our X-ray emission modeling with RIAFs.
In contrast, for the jet model, the gamma rays can be dominated by EIC scattering from relativistic electrons and RIAF target photons.
Neutrino emission in cases where the observed gamma rays are leptonic rather than hadronic has been extensively studied in the context of jet-loud AGNs, including detailed effects of external target photons~\citep{Murase:2014foa,Dermer:2014vaa}. In such jet models, the predicted neutrino fluxes are typically lower than the observed TeV gamma-ray flux, i.e., $L_\nu \lesssim L_\gamma$. This is indeed the case for NGC 4278, and the expected neutrino flux may be 3-4 orders of magnitude lower than the TeV gamma-ray flux \citep[see also][]{chen2026physicaloriginveryhighenergygamma}. Therefore, our neutrino flux can be treated as an upper limit of the two-zone picture consisting of the RIAF and jet in which TeV gamma rays have a leptonic origin. 

Coronal regions of radiatively efficient disks could be regarded as RIAF-like regions, and coronae of Seyfert galaxies have been proposed as the promising origin of medium-energy neutrinos in the 10-100~TeV range~\citep{Murase:2019vdl,Murase:2026}. This model predicts that NGC 1068 and NGC 4151 are among the brightest neutrino sources in the IceCube sky~\citep{Kheirandish:2021wkm,Murase:2023ccp}, as well as a correlation between the neutrino and X-ray luminosities, i.e., an $L_\nu-L_X$ relation. A similar relation is expected in the RIAF model~\citep{Kimura2019}, which tempts us to extend the $L_\nu-L_X$ relation down to lower luminosities, i.e., from Seyfert galaxies to low-luminosity AGNs. 

\begin{figure}[ht]
    \centering
\includegraphics[width=0.75\linewidth]{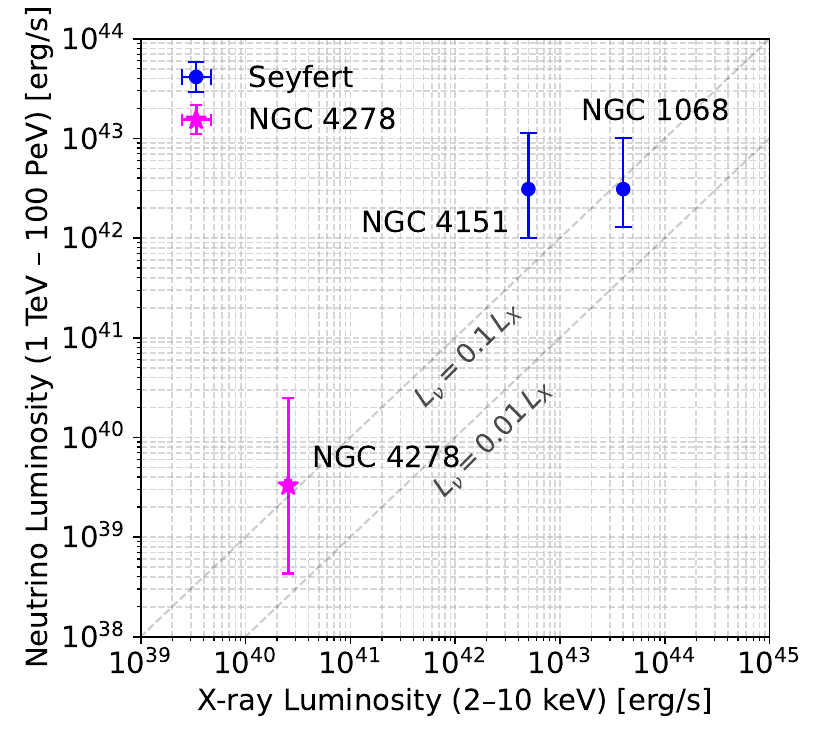}\\
(a)\\

\vspace{1ex}

\includegraphics[width=0.75\linewidth]{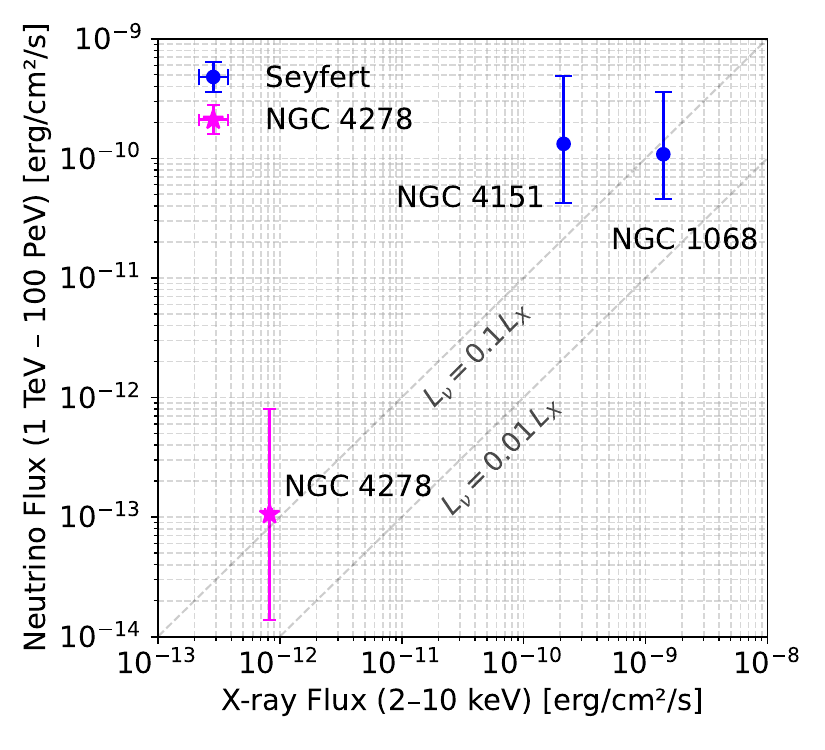}\\
(b)\\
    \caption{%
    (a) Possible scaling relation between the hard X-ray and the expected neutrino luminosities. The data from Seyfert galaxies are from \cite{Carpio2026}. We highlight the caveat that no evidence for any neutrino emission from NGC 4278 has ever been reported. %The hypothetical neutrino flux from NGC 4278 is estimated assuming the entire TeV gamma-ray emission comes from a hadronic origin without internal absorption. 
    (b)~Similar to~(a), but for fluxes.}
\label{fig:xnu}
\end{figure}

Fig.~\ref{fig:xnu}(a) shows a possible scaling relation between the intrinsic 2-10 keV hard X-ray luminosity and the neutrino luminosity from Seyfert galaxies. {\revision We adopt a luminosity distance of 16.1 Mpc \citep{Tonry2001} to convert between flux and luminosity.} The estimation of neutrino luminosities is sensitive to the templates of neutrino spectra, and the power-law assumption is not valid in general, or at least can be misleading for the model comparison. Thus, for the neutrino luminosities of NGC 1068 and NGC 4151, we use the estimates by \cite{Carpio2026}, which obtain lower values than those reported in \cite{Kun24}. Using the observed hard X-ray flux of NGC 4278 (see Section~\ref{subsec:nustar}) and the rough estimation of the neutrino flux above, we show that NGC 4278 follows the same hard X-ray/neutrino correlation seen in Seyfert galaxies. Fig.~\ref{fig:xnu}(b) is similar to Fig.~\ref{fig:xnu}(a) but shows a relation between the hard X-ray and neutrino fluxes. {\revision The dashed lines $L_\nu=0.1 L_X$ and $L_\nu=0.01 L_X$ bracket a physically relevant range for the ratio of the neutrino luminosity to the intrinsic X-ray luminosity, with the lower line being comparable to conservative RIAF models with $P_{\rm CR}/P_{\rm th}\sim0.01$ \citep{Kimura2021} and the upper line representing scenarios with greater cosmic-ray loading such as those explored here. Although the data point for NGC 4278 in Figure \ref{fig:xnu} corresponds to the quiescent state, the scaling relation should also hold for the moderate state since, in our model, the increase in the X-ray flux would result in a proportional increase in the predicted neutrino flux while maintaining the same power law relation. Thus, the data point would shift upward following a line parallel to the dashed lines. It should be noted here that we do not consider the cosmic-ray feedback, so our predictions are best suited for $P_{\rm CR}/P_{\rm th}\lesssim0.1-0.5$. The cosmic-ray pressures are unlikely to be too large, otherwise the dynamical structure of the RIAF would be significantly affected \citep{Kimura:2014rja} or the virial equilibrium would not hold~\citep{Murase:2020lnu}.} %The cosmic rays may no longer behave as passive test particles and can feed back on the MHD turbulence, changing particle acceleration dynamics \citep{Kimura2021,Lemoine:2023wsw}. 
%Therefore, in this regime, our predicted neutrino luminosities should be interpreted as optimistic upper limits.} 
%With a luminosity 2--4 orders of magnitude lower than Seyfert galaxies that exhibit a hard X-ray–neutrino correlation, NGC 4278 potentially extends this correlation down to the LLAGN regime, the most numerous subclass of AGNs.

%Assuming a power law of an index $2$ and a normalization of $2\times10^{-12}\; \mathrm{TeV^{-1}\,cm^{-2}\,s^{-1}}$ at 1 TeV for the neutrino spectrum, we calculate the total neutrino counts expected in the IceCube Neutrino Observatory to be $\sim$0.2 events for the entire 140-day active state, adopting the effective area of IceCube from \cite{2019ICRC...36.1021B}. Similarly, assuming the quiet state (normalization of $3\times10^{-13}\; \mathrm{TeV^{-1}\,cm^{-2}\,s^{-1}}$ at 1 TeV) is representative over a long period, we estimate that roughly 0.7 events would be detected by IceCube over the 10 years. It is, therefore, unlikely that IceCube would detect neutrinos from NGC 4278 even under the optimistic scenario. 

%Using the hypothetical neutrino spectra above, we estimate the 0.3 - 100 TeV neutrino fluxes for the quiet and active states to be $5\times10^{-13}\; \mathrm{erg\,cm^{-2}\,s^{-1}}$ and $2\times10^{-12}\; \mathrm{erg\,cm^{-2}\,s^{-1}}$, respectively. Since there is no detection of neutrinos from NGC 4278 by any experiments, the neutrino flux could be zero. Therefore, we assume the neutrino flux associated with a hadronic TeV gamma-ray origin to be $5^{+15}_{-5}\times10^{-13}\; \mathrm{erg\,cm^{-2}\,s^{-1}}$. 

%%%%%%%%%%%%%%%%%%
%%%%%%%%%%%%%%%%%%
%%%% %     Summary     %%%%
%%%%%%%%%%%%%%%%%%
%%%%%%%%%%%%%%%%%%
\section{Summary}
In this study, we reported recent \swift{}-XRT and \nustar{} X-ray observations for the low-luminosity active galactic nucleus (LLAGN) NGC 4278. The \nustar{} observations led to the first measurements of the hard X-ray spectrum of this source beyond 10 keV, showing a rather flat spectrum with a power-law index around $2$ with no sign of a cutoff toward high energies. These observations reveal a lower-flux state compared to the LHAASO-reported active state. Flux variability by a factor of $\sim2$ was observed on timescales of a month. 

We interpreted the spectral variability of the X-ray emission using a radiatively inefficient accretion flow (RIAF) model. The X-ray variability can be explained by changes in the accretion rate. We found the best-fit parameters for both the quiescent and moderate states. We also explored both single-zone and multi-zone models and concluded that the contribution from the outer layers is negligible in comparison to that from the innermost region. We also found that the magnetically arrested disk is favored within our RIAF model, which is consistent with the jet model for TeV gamma rays observed by LHAASO~\citep{chen2026physicaloriginveryhighenergygamma}.  

The TeV gamma rays observed in LHAASO, on the other hand, cannot be produced in the innermost region of the RIAF disk, and are more likely to come from outer regions {\revision ($R\gtrsim300 R_S$)}, such as the jet or the wind. We studied NGC 4278 as a hidden neutrino source, and LLAGNs have been discussed among the most promising candidates of hidden cosmic-ray accelerators~\citep{Murase:2015xka}.    
We also discussed a possible correlation between the observed hard X-ray and the hypothetical neutrino luminosity that could exist over many orders of magnitude in luminosity between LLAGNs and Seyfert galaxies. 

\begin{acknowledgments}
{\revision We thank the referee for their helpful suggestions.} We thank Jose Carpio for allowing us to show the data points of neutrino luminosities. We also thank Bing Theodore Zhang and Jodi Christiansen for discussions. This research is supported by grants from the U.S. National Science Foundation PHY-2411860 and NASA 80NSSC25K7694. The work of K.M. was supported by the NSF Grants No.~AST-2108466, No.~AST-2108467, and No.~2308021.
S.S.K. acknowledges support by KAKENHI Nos. 22K14028, 21H04487, 23H04899, and the Tohoku Initiative for Fostering Global Researchers for Interdisciplinary Sciences (TI-FRIS) of MEXT's Strategic Professional Development Program for Young Researchers.
This work made use of data supplied by the UK \swift{} Science Data Centre at the University of Leicester.

This research has made use of the NASA/IPAC Extragalactic Database, which is funded by the National Aeronautics and Space Administration and operated by the California Institute of Technology.
\end{acknowledgments}

%% To help institutions obtain information on the effectiveness of their 
%% telescopes the AAS Journals has created a group of keywords for telescope 
%% facilities.
%
%% Following the acknowledgments section, use the following syntax and the
%% \facility{} or \facilities{} macros to list the keywords of facilities used 
%% in the research for the paper.  Each keyword is check against the master 
%% list during copy editing.  Individual instruments can be provided in 
%% parentheses, after the keyword, but they are not verified.

%\vspace{5mm}
\facilities{\nustar{}, \swift{}-XRT} %VLBA}

%% Similar to \facility{}, there is the optional \software command to allow 
%% authors a place to specify which programs were used during the creation of 
%% the manuscript. Authors should list each code and include either a
%% citation or url to the code inside ()s when available.

\software{Astropy \citep{Astropy13,Astropy2018},  
          NumPy \citep{numpy11},
          Matplotlib \citep{Hunter07},
          SciPy \citep{SciPy},
          %VEGAS \citep{Cogan08}, Eventdisplay \citep{Maier17}, Model Analysis \citep[][]{hess_denaurois_2009},
          %Fermitools \citep{Fermitools2019}, 
          HEAsoft \citep{HEAsoft2014}
          }
%% Appendix material should be preceded with a single \appendix command.
%% There should be a \section command for each appendix. Mark appendix
%% subsections with the same markup you use in the main body of the paper.

%% Each Appendix (indicated with \section) will be lettered A, B, C, etc.
%% The equation counter will reset when it encounters the \appendix
%% command and will number appendix equations (A1), (A2), etc. The
%% Figure and Table counter will not reset.

\bibliography{OneBib}{}
\bibliographystyle{aasjournalv7}

%\allauthors

%% This command is needed to show the entire author+affiliation list when
%% the collaboration and author truncation commands are used.  It has to
%% go at the end of the manuscript.
%\allauthors

%% Include this line if you are using the \added, \replaced, \deleted
%% commands to see a summary list of all changes at the end of the article.
%\listofchanges

\end{document}